\documentclass[10pt,aps,prb,twocolumn,longbibliography,superscriptaddress,floatfix]{revtex4-2}

\usepackage{graphicx} % Include figure files
\usepackage{bm}% bold math
\usepackage{color}
\usepackage[table]{xcolor}
\usepackage{makecell}
\usepackage{multirow}
\usepackage{makecell}
\usepackage{tabularx}
\usepackage{subfigure}

\usepackage{amsmath}
\usepackage{contour}
\usepackage[normalem]{ulem}
\usepackage{comment}
\usepackage{amssymb}
\usepackage[percent]{overpic}
\usepackage{comment}
\usepackage{bbold}
\usepackage{enumerate}
\usepackage{float}
\usepackage{xspace}
\usepackage{mathtools}
\usepackage{mathrsfs}
\usepackage{mathptmx}
\usepackage[colorlinks=true,linkcolor=blue,citecolor=blue,urlcolor=blue]{hyperref}
\usepackage{orcidlink}
\usepackage{tikz}
\usepackage{siunitx}
\usepackage[utf8]{inputenc}
\usepackage{nicefrac}
\usepackage[capitalise]{cleveref}

\begin{document}
\title{Thickness-dependent magnon spin transport in antiferromagnetic insulators: Crossover from quasi-three-dimensional to quasi-two-dimensional regime}

\author{Mathias Åsan Myhre} 
\altaffiliation{These authors contributed equally to this work.}
\affiliation{Center for Quantum Spintronics, Norwegian University of Science and Technology, 7034 Trondheim, Norway} 
\author{Verena Brehm \orcidlink{0000-0002-7174-1899}}
\altaffiliation{These authors contributed equally to this work.}
\email{v.j.brehm@tue.nl}
\affiliation{Center for Quantum Spintronics, Norwegian University of Science and Technology, 7034 Trondheim, Norway}
\affiliation{Department of Applied Physics and Science Education,
Eindhoven University of Technology, 5612 AP Eindhoven, Netherlands}
\author{Thomas Delvaux \orcidlink{0009-0006-9797-7425}}
\affiliation{ENSTA, Institut Polytechnique de Paris, 828, Boulevard des Maréchaux, 91762 Palaiseau Cedex, Paris, France}
\author{Arne Brataas \orcidlink{https://orcid.org/0000-0003-0867-6323}} 
\affiliation{Center for Quantum Spintronics, Norwegian University of Science and Technology, 7034 Trondheim, Norway}
\author{Alireza Qaiumzadeh \orcidlink{https://orcid.org/0000-0003-2412-0296}} 
\affiliation{Center for Quantum Spintronics, Norwegian University of Science and Technology, 7034 Trondheim, Norway}

\begin{abstract}
Motivated by the recent observation of giant room-temperature magnon spin conductivity in an ultrathin ferromagnetic insulator [X.-Y. Wei \emph{et al.},  \href{https://doi.org/10.1038/s41563-022-01369-0} {Nat. Mater. $\bm{21}$, 1352 (2022)}], we investigate thickness-dependent magnon spin transport in thin antiferromagnetic insulators (AFIs). We study the prototypical AFI hematite, known for its exceptionally low magnetic damping and two distinct magnetic phases: a low-temperature uniaxial easy-axis phase and a high-temperature biaxial easy-plane phase. Using stochastic micromagnetic simulations, we investigate thickness-dependent magnon spin transport across both magnetic phases. Our results uncover a crossover from quasi-three-dimensional to quasi-two-dimensional magnon spin transport at a critical thickness determined by the frequency or energy of the excited magnons. Below this critical thickness, we observe a pronounced enhancement in the magnon diffusion length in both magnetic phases. This rise is attributed to a change in the effective magnon density of states, reflecting the reduced phase space available for scattering in the thinner, quasi-two-dimensional regime. Understanding and controlling long-distance magnon spin transport in AFIs is crucial for developing next-generation spintronic nanodevices, especially as materials approach the two-dimensional limit.
\end{abstract}

\maketitle

\section{Introduction}
Modern spintronics relies on the control and long-distance transport of spin angular momentum in magnetic materials. Due to their avoidance of Joule heating and abundance, electrically insulating magnetic materials are especially promising \cite{spin_insulatronics}. Long-distance magnon spin transport has recently been observed in both ferromagnetic insulators (FMIs) and antiferromagnetic insulators (AFIs)\cite{long_distance_roomT,tunableLongDistanceKlaui}. 
    
AFIs are especially promising for spintronic applications due to their negligible stray fields, ultrafast THz-range spin dynamics, and the presence of two degenerate magnon modes carrying opposite spin angular momentum \cite{AFMreviewManchon,AFMreviewHelen}. 
One prominent system for spin transport in AFI is hematite $\alpha$-Fe$_2$O$_3$, which offers not only exceptionally low damping but also two magnetically ordered phases: a uniaxial easy-axis collinear phase below the Morin temperature and a biaxial easy-plane canted phase above it \cite{MorrishHematite}. 
The diffusive transport of magnon spin across micrometer distances has been demonstrated experimentally using a nonlocal geometry in both the easy-axis collinear phase \cite{tunableLongDistanceKlaui,2020NanoL..20..306R} and the canted easy-plane phase  \cite{tunableLongDistanceKlaui,han_birefringence-like_2020,2020ApPhL.117x2405R,2022JMMM..54368631R,akash1}. 
In the easy-axis magnetic phase, hematite supports two degenerate, circularly polarized magnon modes, each carrying finite and opposite spin angular momentum. In contrast, in the easy-plane magnetic phase, the magnon eigenmodes become linearly polarized and, as a result, do not carry any net spin angular momentum individually.
This makes the experimentally observed finite spin angular momentum transport in the easy-plane magnetic phase of hematite particularly intriguing. In Refs. \cite{tunableLongDistanceKlaui,han_birefringence-like_2020}, this phenomenon was attributed to a birefringence-like mechanism. Alternatively, Refs. \cite{akash1,akash2} proposed a phenomenological theory based on the dynamics of the magnon pseudospin, where a finite homogeneous Dzyaloshinskii–Moriya (DM) interaction is considered essential.
However, more recent theoretical studies and micromagnetic simulations  have demonstrated that the observed Hanle-like magnon spin transport can be attributed to intermode coherent magnon beating between two orthogonal and linearly polarized magnon eigenmodes. This mechanism persists even in the absence of DM interaction, establishing it as a generic characteristic of easy-plane AFIs
\cite{verena_micromagnetic_study_transport_AFM,BeatingTheoryYaroslav}. 

    \begin{figure}
        \centering
        \includegraphics[width=\linewidth]{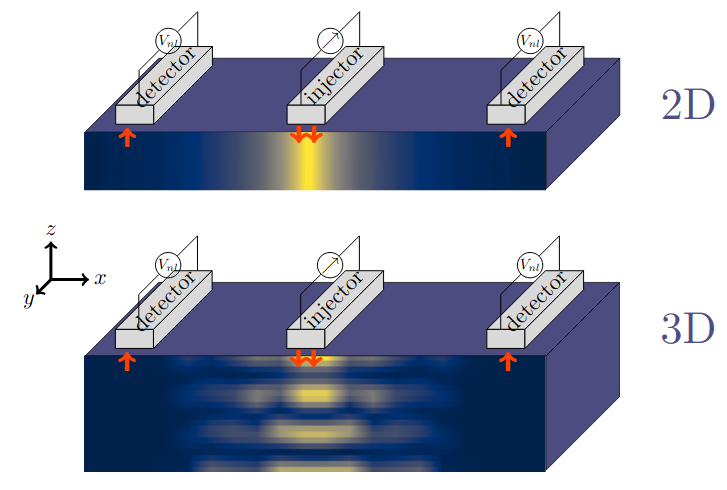}
        
        \caption{Setup for micromagnetic simulations of magnon spin transport in a nonlocal transverse geometry. Magnons are injected into the AFI from the central lead via interfacial torque generated by a spin Hall–induced spin accumulation. This increases the magnon spin chemical potential, driving diffusive transport toward the left and right leads along the $x$ direction. Through spin pumping and the inverse spin Hall effect, the magnon spin signal is detected at various distances. The spatial magnon distribution along the thickness show a clear difference between quasi-2D (top) and quasi-3D (bottom) magnon spin transport regimes.}
        \label{fig:transport_setup_schematic}
    \end{figure}
    
While most magnon spin transport experiments in both FMIs and AFIs have traditionally been performed on three-dimensional (3D) bulk crystals, recent experimental advances in film growth and detection techniques have enabled a shift toward studying thinner films and two-dimensional (2D) magnetic systems. 2D magnetic systems \cite{RevModPhys.92.021003,Mounet2018,Gibertini2019,doi:10.1126/science.aav4450} represent a novel and promising platform for the next generation of magnonic and spintronics \cite{Chumak2015,Cornelissen2015LongDistSpinTransp,MagSpinTransportChemPot} and spintronic nanodevices \cite{Kaverzin_2022,2DspintronicsReview,MagneticGenom2DvdW2022}. For example, the fashionable magnetic van der Waals (vdW) materials, which provide atomically thin layers to study 2D magnetism \cite{doi:10.1126/science.aav4450,Magnetism_in_2D_van_der_Waals_materials,RamanFePs3Wang2016,Bedoya_Pinto_2021,Huang2017}, have gained a great deal of attention. 
Recent advances in material synthesis have extended beyond vdW materials \cite{nihei2025nonvanderwaalsheterostructures}. Notably, liquid exfoliation techniques applied to non-vdW crystals have enabled the production of 2D hematite nanosheets, coined as hematene \cite{puthirath2018exfoliation}.

The study of magnon spin transport in thin films and 2D magnetic layers is still in its infancy, with many fundamental mechanisms yet to be fully understood \cite{PhysRevX.9.011026,LongDistMagTransportWees,PhysRevB.110.174440,PhysRevResearch.4.043214, verenaCrCl,VerenaCrI3Nernst}.
Recent observations have revealed exceptionally large magnon spin conductivity in ultrathin Yttrium Iron Garnet (YIG) films, a prototype FMI, with a pronounced enhancement over thicker films at room temperature. The observed spin conductivity rivals that of high-mobility 2D electron gases in GaAs quantum wells at millikelvin temperatures \cite{GiantSpinConductivityYIG2022,jungfleisch2022two}, paving the way for low-dissipation magnonic nanodevices. 
    
Inspired by this observation, we investigate whether a similar phenomenon may occur in AFIs and aim to uncover the underlying mechanism. To address these questions, we perform finite-temperature micromagnetic simulations to study how magnon spin transport in hematite depends on film thickness in both its easy-axis and easy-plane magnetic phases. Our results reveal a pronounced enhancement in the magnon diffusion length at a critical thickness, which we interpret as a transition from quasi-3D to quasi-2D transport. We justify this finding based on the density of states (DOS).

The article is organized as follows. After introducing our model in Sec. \ref{sec:model}, we will first show the thickness-dependent transport signals in Sec. \ref{sec:transportSignals}, along with the extracted spin diffusion lengths. Here, the transition from quasi-2D to quasi-3D transport becomes evident. In Sec. \ref{sec:Discussion} we discuss our findings along with analytical arguments based on the magnon density of states. Finally, we conclude in Sec. \ref{sec:Conclusion}.

\section{System Model} \label{sec:model}
In this section, we present the system setup, the underlying physical model, simulation methodology, and the magnon dispersions and DOS in our two magnetic models.

\subsection{Nonlocal geometry for magnon spin transport}
An experimental transverse nonlocal geometry \cite{tunableLongDistanceKlaui} can be emulated with a computational setup as schematically shown in \cref{fig:transport_setup_schematic}. Magnons are injected through the central lead by interfacial spin-transfer torque, driven by a spin accumulation generated via the spin Hall effect in the heavy-metal injector. This local rise in the nonequilibrium magnon chemical potential \cite{FlebusMagChemPot} drives magnon diffusion across the AFI layer toward the left and right leads. The transported spin angular momentum is detected through spin pumping from the AFI layer into the adjacent heavy-metal detector, where it is converted into a measurable electric voltage via the inverse spin Hall effect.

\subsection{Micromagnetic simulations}
Our model for the easy-axis phase is based on the low-temperature easy-axis magnetic phase of hematite, in which two circularly polarized magnon modes are degenerate in the absence of an external magnetic field. For the easy-plane phase, which represents the high-temperature phase of hematite, we follow the approach of Ref.~\cite{verena_micromagnetic_study_transport_AFM}, incorporating a transverse critical magnetic field. In this regime, the two linearly polarized  magnon bands become degenerate, and magnon spin transport exhibits conventional exponential decay. 
Furthermore, we ignore canting induced by DM interaction in the easy-plane phase, since it was shown in Ref. \cite{verena_micromagnetic_study_transport_AFM} that the impact on the transport is negligible.      

In order to simulate the magnetization dynamics of these two magnetic phases of hematite, we conduct micromagnetic simulations at finite temperature using the open-source code \textsc{BORIS} \cite{BORIS}. Every discrete simulation cell with a small volume $V$ is assigned a macrospin magnetic moment $\bm{M}_i$ with a homogeneous saturation magnetization $M_s$ \cite{AtomMicroScales}. 
The dynamics of the magnetic moment direction $\bm{m}_i = \bm{M}_i/M_s$ are described by coupled stochastic Landau-Lifshitz-Gilbert (sLLG) equations,
\begin{align}
        \dot{\bm{m}}_i = &- \gamma \bm{m}_i \times \left(\bm{H}_i^\text{eff} + \bm{H}^\text{th}_i\right) \nonumber \\ & - \alpha \gamma \bm{m}_i \times \left[ \bm{m}_i \times \left(\bm{H}_i^\text{eff} + \bm{H}^\text{th}_i\right)\right].
\end{align}
Here, $i \in \{A, B\}$ refers to the two antiferromagnetic (AF) sublattices. We define $\gamma \equiv \mu_0 |\gamma_e| / (1 + \alpha^2)$, where $\mu_0$ is the vacuum permeability, $\gamma_e = -g \mu_B / \hbar$ is the electron gyromagnetic ratio, $g$ is the $g$-factor, $\mu_B$ is the Bohr magneton, $\hbar$ is the reduced Planck constant, and $\alpha$ is the dimensionless Gilbert damping parameter \cite{ASDeriksson}.
$\bm{H}_i^\text{eff}$ is the effective magnetic field at site $i$, and $\bm{H}_i^\text{th}$ is a stochastic thermal field that adds temperature to the model. 
A normalized Gaussian distribution is scaled with the prefactor
$\xi_{th}= \sqrt{{2\alpha k_B T}/{\big(\gamma \mu_0 M_s V \Delta t\big)}}$ in every component, adding white noise to the system that is weighted with the thermal energy $k_B T$. Here, $k_B$ represents the Boltzmann constant, and it is scaled with both the cell volume $V$ and the time step of the simulation  $\Delta t$. 
    
In our effective spin model for hematite, the total effective magnetic field includes contributions from exchange interactions, magnetic anisotropies, the external magnetic field, and spin-Hall-induced spin–orbit torque.
\begin{equation}
    \bm{H}_i^\text{eff} = \bm{H}_i^\text{ex}  + \bm{H}_i^\text{ani} + \bm{H}_i^\text{ext} + \bm{H}_i^\text{SOT}. 
    \end{equation} 
$\bm{H}_i^\text{ex}$ is the sum of homogeneous and inhomogeneous AF exchange interactions \cite{BORIS},
\begin{align}
        \bm{H}_{i}^\text{ex} = - \frac{4 A_h}{\mu_0 M_s} \left[\bm{m}_i \times \left(\bm{m}_i \times \bm{m}_j \right)\right]+\frac{2 A}{\mu_0 M_s} \nabla^2 \bm{m}_i 
\end{align} 
where $i \neq j$, $A_h$ are the homogeneous AF exchange constants, and $A$ is the AF exchange stiffness.

$\bm{H}_i^\text{ani}$ denotes the magnetic anisotropy fields \cite{BORIS},
\begin{align}
        \bm{H}_i^\text{ani} = \bm{H}_i^\text{hard} + \bm{H}_i^\text{easy} = \sum_{l \in \{\text{hard},\text{easy}\}} \frac{2 K_l}{\mu_0 M_s} \left(\bm{m}_i \cdot \hat{e}_l\right) \hat{e}_l,
\end{align} 
where the hard-axis anisotropy is given by $K_\text{hard} > 0$ and $\hat{e}_\text{hard} = \hat{z}$, and the easy-axis anisotropy is along $\hat{e}_\text{easy}=\hat{x}$ with $K_\text{easy} < 0$. This is the only term that distinguishes between the two hematite magnetic phases: the easy-axis phase includes only easy-axis magnetic anisotropy, which lies along the $x$ axis, while the easy-plane phase features one strong hard-axis magnetic anisotropy along the $z$ direction and an additional weak easy-axis anisotropy along the $x$ direction. In both cases, the magnetic ground state is collinear and aligned along the easy axis.

The term $\bm{H}_i^\text{ext}$ denotes the external magnetic field, which couples to the AF moments via the Zeeman interaction. This field is only applied in the easy-plane magnetic phase, oriented perpendicular to both the easy-axis and hard-axis anisotropy directions, specifically along the $y$ axis. As shown in Ref.~\cite{verena_micromagnetic_study_transport_AFM}, applying an external field along this direction with a critical strength of approximately \SI{6}{T} induces the degeneracy of the two linearly polarized magnon branches. Consequently, spin transport becomes diffusive, characterized by an exponential decay of the spin accumulation.

Finally, $\bm{H}_i^\text{SOT}$ is the total spin-orbit torque (SOT), which is the sum of a field-like and an antidamping-like torque \cite{BORIS},
    \begin{align} \label{eq:torque}
        \bm{H}_i^\text{SOT} = - \frac{\Theta}{\gamma M_s}  \frac{\mu_B}{e}\frac{|J_c|}{L_z}\left(\bm{m}_i \times \bm{P} + r_G \bm{P} \right),
    \end{align} 
generated by a DC charge current with the density $J_c$, which is converted to a spin current via the spin Hall effect. $\bm{P}$ is the direction of spin-Hall-induced spin polarization at the interface. Furthermore, $\Theta$ is the spin Hall angle, a measure of the efficiency of charge-to-spin current conversion; $r_G$ parameterizes the field-like torque amplitude, and $L_z$ is the thickness of the AFI layer.
In order to model the spin-Hall-induced SOT, see \cref{fig:transport_setup_schematic}, we set $\Theta > 0$ in the injector region
and $\Theta=0$ otherwise. The direction of $\bm{P}$ lies along the easy-axis direction, so there is no excitation at zero temperature since $\bm{m}_i\parallel \bm{P}$. Finite temperature, however, induces thermal fluctuations in magnetic moments $\bm{m}_i$, and therefore, the net spin torque is finite; consequently, magnons are pumped into the underlying AFI layer. For more details concerning this technique, please refer to \cite{verena_phd,verena_micromagnetic_study_transport_AFM}. 

We neglect the homogeneous DM and dipolar interactions in our hematite model, as they do not play a significant role in magnon spin transport \cite{verena_micromagnetic_study_transport_AFM}.

The magnon spin current is detected at the heavy-metal detector via the inverse spin Hall effect. The resulting inverse spin Hall effect voltage is proportional to the spin current injected into the detector, which, in turn, is driven by the spin accumulation at the interface between the AFI and the detector, given by \cite{ArneSpinPumping, PhysRevB.102.020408},
    \begin{equation} \label{eq:spinAccum}
        \bm{\mu}({x}) := G_r^{\uparrow \downarrow}  \left< \sum_{i}\big[\bm{m}_i(t,x)\times \dot{\bm{m}}_i(t,x)\big]\right>,
    \end{equation}
where $G_r^{\uparrow \downarrow}$ %  
is the real part of the spin mixing conductance \cite{ArneSpinMixingConductance}, $\left< \cdot \right>$ denotes an average, and $x$ is the distance between the injector and detector. 
The time average for each ensemble member starts after the steady state is reached.
In our setup geometry, \cref{fig:transport_setup_schematic}, the inverse spin Hall detector measures only the $x$ component of the spin accumulation at the interface; we define $\mu := \mu_x(x)$.

The effective thickness of each layer is determined by the size of our micromagnetic simulation cell, which is set to \SI{5}{nm} in our simulations. Other parameters used in our simulations are listed in \cref{tab:system_parameters}. To investigate the 2D–3D crossover, we examine samples with thicknesses ranging from a single layer up to eight layers (\SI{40}{nm}). This maximum thickness remains much smaller than the system’s diffusion length, which is on the order of \SIrange{0.1}{0.5}{\micro\meter}. Consequently, magnon decay along the thickness is negligible in our model.

\subsection{Magnon dispersion and density of states} \label{sec:DOScalculation}
In the continuum limit, the bulk magnon dispersion of an AFI with easy-axis magnetic anisotropy is given by \cite{verena_phd},
\begin{equation}\label{eq:uniaxialDispersion}
        E_{uni}^{\pm}(q) = C\sqrt{|A_{h}||K_{easy}| + A|A_h| q^2},
\end{equation}
where $C \equiv \frac{4\hbar \gamma_e}{M_s}$, and the $\pm$ sign corresponds to the two circular magnon modes with opposite polarization. The two magnon modes are degenerate in this case.

In the easy-plane phase, however, the two magnon eigenmodes are linearly polarized and are split by a finite gap. The dispersion relations for the two orthogonal linearly polarized modes are given by
\cite{verena_phd}
\begin{align}\label{eq:biaxialDispersion}
    E_{bi}^+(q) &= C\sqrt{A|A_h|q^2 + |A_h|(|K_{easy}|+K_{hard})}, \\
    E_{bi}^-(q) &= C\sqrt{A|A_h|q^2 + |A_h||K_{easy}| + \left(\frac{1}{4}M_s\mu_oH_y\right)^2}.
\end{align}
Note that, even in the absence of a magnetic field, these two magnon modes are not degenerate. At a critical transverse field of $H^c_y=\pm  4\sqrt{|A_h|K_{hard}}/M_s\mu_o$, however, the degeneracy can be restored.

We are interested in the DOS because it determines the number of quasiparticle modes available to participate in transport at a given energy. The DOS depends on both the system dimensionality and the quasiparticle dispersion and is defined as
\begin{equation}
        g_D(E) = \int \frac{\text{d}^D \bm{q}}{(2\pi)^D}\delta\big(E-E(q)\big),
\end{equation}
where $D$ is the spatial dimension and $\delta(...)$ is the Dirac delta function. For magnons in 2D and 3D AFIs, we find 
\begin{align}\label{eq:DoS_AFM_2D}
    g_2^{AF}(E) &= \frac{|E|}{2\pi C^2A|A_h|}\Theta\big(E^2-E_0^2\big), \\ 
\label{eq:DoS_AFM_3D}
    g_3^{AF}(E) &= \frac{|E|\sqrt{E^2-E_0^2}}{2\pi^2 C^3(A|A_h|)^{3/2}}\Theta\big(E^2-E_0^2\big).
\end{align}
where $\Theta(...)$ is the Heaviside step function and $E_0$ is the magnon gap. In the easy-axis magnetic phase, $E_0 = C\sqrt{|A_h| K_{easy}}$ is the degenerate magnon gap, while in the easy-plane magnetic phase, $E_0^+ = C\sqrt{|A_h| (K_{easy}+K_{hard})}$ and $E_0^- = C\sqrt{|A_h| K_{easy}+\left(M_s\mu_0 H_y/4 \right)^2}$ are the magnon gaps of the upper and lower branches, respectively. However, as we have already mentioned, by applying the critical transverse magnetic field, two linearly polarized magnon modes are tuned to be degenerate so that $E_0^- = E_0^+$. 

It is worth noting that in AF systems, the DOS vanishes at zero energy with different power laws in 2D and 3D; see Eqs. (\ref{eq:DoS_AFM_2D}) and (\ref{eq:DoS_AFM_3D}).  
The different DOS behaviors in 2D and 3D reveal that, at low energies, 2D systems can offer more available states than their 3D counterparts. These distinctions play a crucial role in shaping magnon spin transport, as will be elaborated in the following sections.

\section{Magnon spin transport} \label{sec:transportSignals}
\begin{figure}
    \centering
    \includegraphics[width=\linewidth]{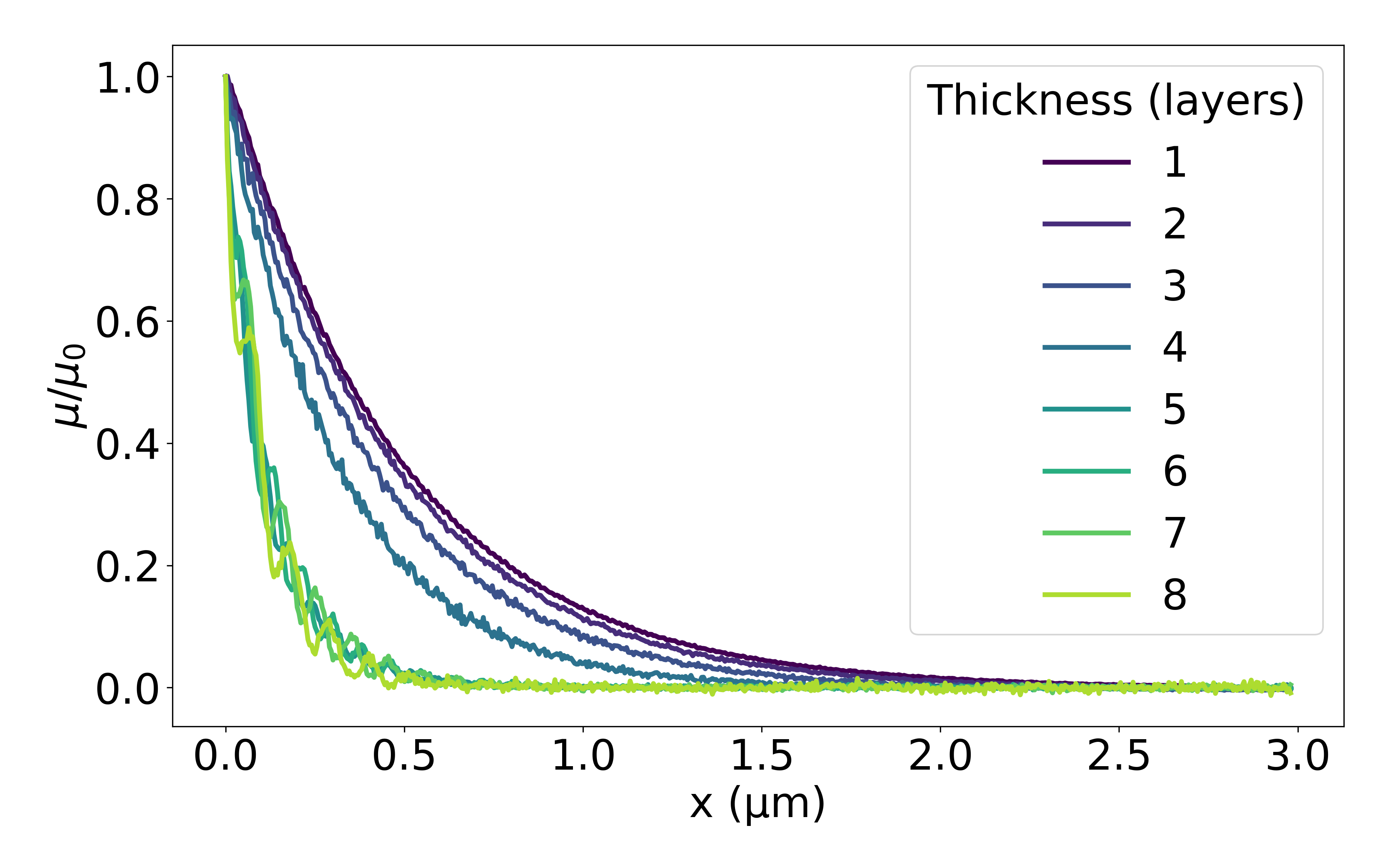}
    \caption{Distance-dependent magnon spin signal in the easy-axis magnetic phase of hematite for varying layer thicknesses. The magnon spin accumulation or magnon chemical potential $\mu$ is normalized by the magnon accumulation at the injector $\mu_0\equiv\mu(x=0)$. The thickness of each layer is \SI{5}{nm} and we set $V/L_z= \SI{12}{\micro\volt\per\nano\meter}$.}
    \label{fig:uniaxial_distance-dependent_transport}
    \end{figure}

In \cref{fig:uniaxial_distance-dependent_transport}, we present the distance-dependent magnon spin signal along the magnetic ground state $\mu(x)$, for various system thicknesses in the easy-axis  magnetic phase. 
In order to facilitate the comparison, we normalize all curves to the spin accumulation at the injector, $\mu_0$. 
Across all thicknesses, the spin signal decays exponentially with distance, consistent with the diffusive magnon spin transport theory \cite{WeesMagnonDiffusion}. Within our chosen spin parameters, systems with four or fewer layers exhibit significantly longer diffusion lengths than those with five or more layers, revealing a clear thickness-dependent crossover. A qualitatively similar behavior is observed in the easy-plane  magnetic phase as well; see \cref{fig:biaxial_distance-dependent_transport}. 
\begin{figure}
        \centering
        \includegraphics[width=\linewidth]{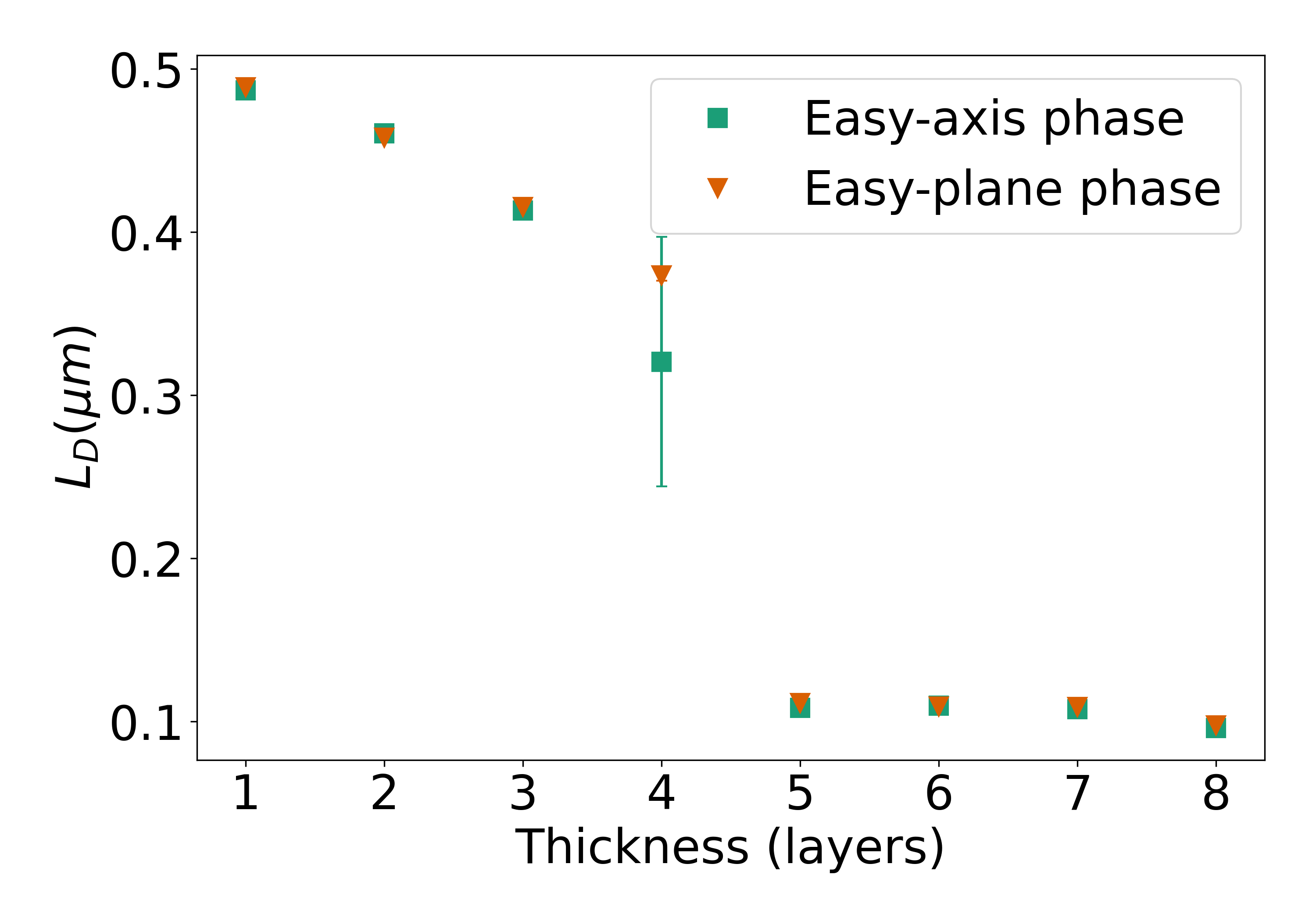}
        \caption{Spin diffusion length for systems in the easy-axis (green squares) and easy-plane (orange triangles) magnetic phases of hematite for varying thicknesses, as obtained by fitting the exponential decay shown in \cref{fig:uniaxial_distance-dependent_transport,fig:biaxial_distance-dependent_transport}. There is a clear enhancement of the diffusion length at a thickness of four to five layers, corresponding to $20$-\SI{25}{nm}, marking the transition from quasi-3D to quasi-2D transport. We set $V/L_z= \SI{12}{\micro\volt\per\nano\meter}$. The large error bar at the critical thickness of four layers suggests a coexistence of 2D and 3D behavior in the easy-axis phase, see \cref{fig:diffusion_length_over_voltage_4layer}.}
\label{fig:diffusion_lengths_across_thicknesses}
\end{figure}
    
To determine the propagation lengths and extract the diffusion length $L_D$, the data are fitted using a single exponential decay of the magnon spin signal $\mu(x) = \mu_0 \exp(-x/L_D)$.
Oscillations superimposed on the exponential decay of the spin signal, seen in the green curves for thicker systems in \cref{fig:uniaxial_distance-dependent_transport}, do not alter our primary conclusion regarding the quasi-2D to quasi-3D crossover. We attribute these oscillations to magnon reflections at the bottom interface and interference with surface modes.
In \cref{sec:remarkOscillations}, we demonstrate that for thicknesses larger than the critical crossover thickness, the spin signal requires fitting with a two-component model.

The thickness-dependent diffusion lengths are shown in \cref{fig:diffusion_lengths_across_thicknesses} with green squares for the easy-axis and orange triangles for the easy-plane magnetic phases. For film thicknesses between four and five layers, corresponding to $20$–\SI{25}{nm}, both magnetic phases show a pronounced enhancement in the thinner films. In the thinner films, below the crossover thickness, the diffusion length is approximately four times larger than that in the thicker films. As we argue in the following section, we interpret this critical crossover thickness as the transition from a quasi-3D to a quasi-2D transport regime. Within our model framework, we find that, at the critical thickness, the diffusion length in the easy-axis phase shows substantial variation, indicating enhanced sensitivity of transport properties near the dimensional crossover, see \cref{fig:diffusion_length_over_voltage_4layer}.  This is visualized in by the large error bar.

To verify whether the observed critical thickness arises from the finite system width or the chosen spin-torque amplitudes, we performed a series of systematic numerical tests, varying both the system width and injection strength. The critical thickness was found to be largely independent of either parameter; see \cref{sec:ImpactWidthAndTorqueStrength}.

\section{Discussion} \label{sec:Discussion}
In the previous section, we demonstrated that the diffusion length, and hence spin conductivity, for magnon spin transport in both easy-axis and easy-plane AFIs depends strongly on system thickness. There appears to be a critical thickness beyond which the magnon diffusion length drops to approximately 20\% of its value in a single-layer system. 

In Fig.~\ref{fig:transport_setup_schematic}, we show the spatial magnon profile along the thickness for thin and thick systems under comparable effective spin-torque amplitudes. It is evident that, while the magnon profile remains nearly uniform in the thin limit, it becomes spatially modulated along the thickness in thicker films. This suggests that the magnon wavenumber may play a key role in governing the transition from the 2D to the 3D transport regime.

 \begin{figure}
        \centering
        \includegraphics[width=\linewidth]{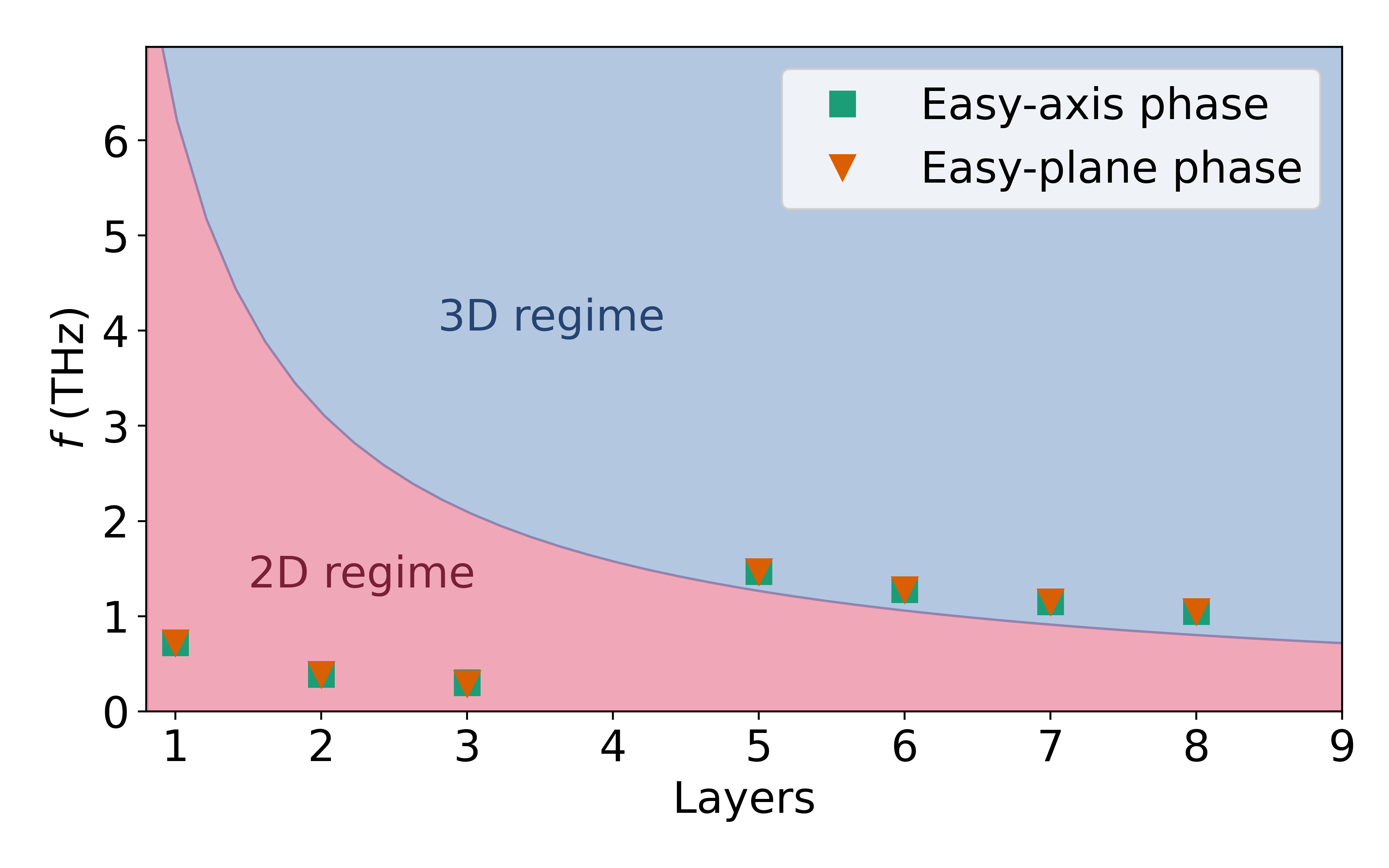}
        \caption{Measured frequencies of excited magnon modes during the transport simulations in systems in the easy-axis (green squares) and easy-plane (orange triangles) magnetic phase for varying thickness. The frequency is read out from spectra that are recorded during the transport experiments, see \cref{sec:ReadOutOccupiedStates} for more details. The colored background shows analytically calculated regions of the 2D (red area) versus 3D (blue area) regime. The gray line shows the critical frequency dividing the two transport regimes, see \cref{eq:criticalE}. The data points corresponding to the critical thickness (four layers) are not shown, as the occupied frequencies exhibit substantial variation at this transition point.}
        \label{fig:critical_frequencies_with_measurements}
    \end{figure}

By analyzing the magnon DOS in both 2D and 3D, we identify a characteristic magnon frequency or energy at which the system transitions from a quasi-2D to a quasi-3D regime, depending on the film thickness. Given that the 2D and 3D DOS have units of $\mathrm{J}^{-1} \, \mathrm{m}^{-2}$ and $\mathrm{J}^{-1} \, \mathrm{m}^{-3}$, respectively, the ratio between them defines a characteristic length scale, interpreted as the critical thickness separating the two transport regimes.
Accordingly, we expect magnons with frequencies below the critical frequency to exhibit quasi-2D transport behavior, while magnons with higher frequencies, above the critical frequency, are expected to contribute to quasi-3D transport. The critical magnon energy or frequency at each thickness $L_z$ can be found via, 
\begin{equation}\label{eq:critical_frequencies}
        \frac{g_2^{AF}}{g_3^{AF}} = \frac{\pi C\sqrt{A|A_h|}}{ \sqrt{E_c^2-E_0^2}} = L_z,
    \end{equation}
and thus
    \begin{equation} \label{eq:criticalE}
        |E_c| = \sqrt{\pi^2L_z^2C^2A|A_h| + E_0^2},
    \end{equation}  
giving us the critical magnon energy or frequency for a given thickness $L_z$.

Below this critical frequency, at a fixed system thickness, the spin accumulation profile along the thickness, $z$ direction, remains nearly uniform. However, above the critical frequency, standing wave patterns emerge along the $z$ direction. 

Using the analytical expressions for the magnon dispersions, we computed the DOS for both magnetic phases; see Sec.~\ref{sec:DOScalculation}. By identifying the crossing point between the DOS of the 2D system, which scales linearly with the magnon energy, and that of the 3D system, which scales quadratically with the magnon energy, we determine a critical frequency for each thickness that separates quasi-2D-like and quasi-3D-like transport regimes. The analytically found regions are shown with a colored red and blue background in \cref{fig:critical_frequencies_with_measurements}. We then compare these two regimes to the frequency of the excited magnons in our simulations. These are the states excited by interfacial spin torque. In order to find these frequencies, we record a spectrum of occupied magnon states during the transport simulation, using a fast Fourier transform of the spin configuration and reading out the frequencies of the dominantly occupied states. More details are discussed in \cref{sec:ReadOutOccupiedStates}. In \cref{fig:critical_frequencies_with_measurements}, we show these numerically readout magnon excitation frequencies with green squares for the magnetic easy-axis phase and orange triangles for the magnetic easy-plane phase. We observe that these frequencies lie in the 2D regime for thicknesses smaller than the critical thickness and in the 3D regime for thicknesses above the critical thickness. At the crossover thickness (four layers), magnons excited by the spin torque show large frequency fluctuations across different micromagnetic realizations, reflecting both quasi-2D and quasi-3D transport behavior; see Fig. \ref{fig:diffusion_length_over_voltage_4layer}.

The characteristic length scale of the system, the domain-wall length set by the exchange energy compared to the magnetic anisotropy energy, is $\lambda \propto \sqrt{A/2K} \approx \SI{42}{nm}$ for the leading anisotropy term. Therefore, as sketched in \cref{fig:transport_setup_schematic}, below the critical thickness where mostly low-frequency magnons are occupied, the precession along the $z$ direction is quasi-coherent. Above the critical thickness, where magnons with a wider range of energies and wavelengths are excited, the coherence along the $z$ direction is reduced, standing waves can form and interfere, and thus the diffusion length is reduced.

Transport is influenced not only by the DOS but also by the energy distribution of the quasiparticles, which determines how the available states are occupied. In our stochastic micromagnetic simulations, thermal fluctuations are modeled using an uncorrelated white-noise thermostat, which effectively enforces a Rayleigh–Jeans distribution that is valid for low-energy magnons in the high-temperature limit. While magnons are bosonic and should generally follow a Bose–Einstein distribution \cite{weber2025atomisticspindynamicsquantum}, the classical approximation holds for the relevant energy scales and temperatures considered here. Whether the use of a correlated noise thermostat that reproduces the full Bose–Einstein distribution would influence magnon spin transport in AFI thin films remains an open question. 
We suggest performing thickness- and temperature-dependent measurements of the spin diffusion length in various AFIs to uncover the underlying dimensional crossover and the potential role of quantum statistical effects in magnon spin transport.

\section{Summary and Concluding remarks}\label{sec:Conclusion}
Using stochastic micromagnetic simulations, we have demonstrated the strong thickness dependence of magnon spin transport in the AFI hematite, considering both easy-axis and easy-plane magnetic phases. Our results reveal a quasi-3D to quasi-2D crossover at a critical thickness. Below this crossover thickness, the magnon diffusion length sharply increases to about four times its value in the quasi-3D regime. This transition can be attributed to two key factors: (i) the emergence of standing spin-wave modes along the thickness direction, and (ii) the shift in the density of available magnonic states, as captured by DOS, relative to the dominant energy scale of the transported magnons.

These findings underscore the importance of dimensionality in magnon spin transport and highlight ultrathin AFIs as promising platforms for the development of next-generation spintronic devices. 
    
\section*{Acknowledgments}
This work has been supported by the Research Council of Norway through its Centers of Excellence funding scheme, Project No. 262633 ``QuSpin'' and FRIPRO with Project No. 353919 ``QTransMag''. 

\section*{DATA AVAILABILITY}
The simulation scripts that have been used for generating the findings of this article are openly
available \cite{GithubRepo}.
    \appendix

    \renewcommand{\thefigure}{A\arabic{figure}}
    \renewcommand{\thetable}{A\arabic{table}}
    \renewcommand{\thesection}{\Alph{section}}
    \setcounter{figure}{0}  % Reset counter

\section{Magnon spin transport in the easy-plane magnetic phase}  
\begin{figure}
        \centering
        \includegraphics[width=\linewidth]{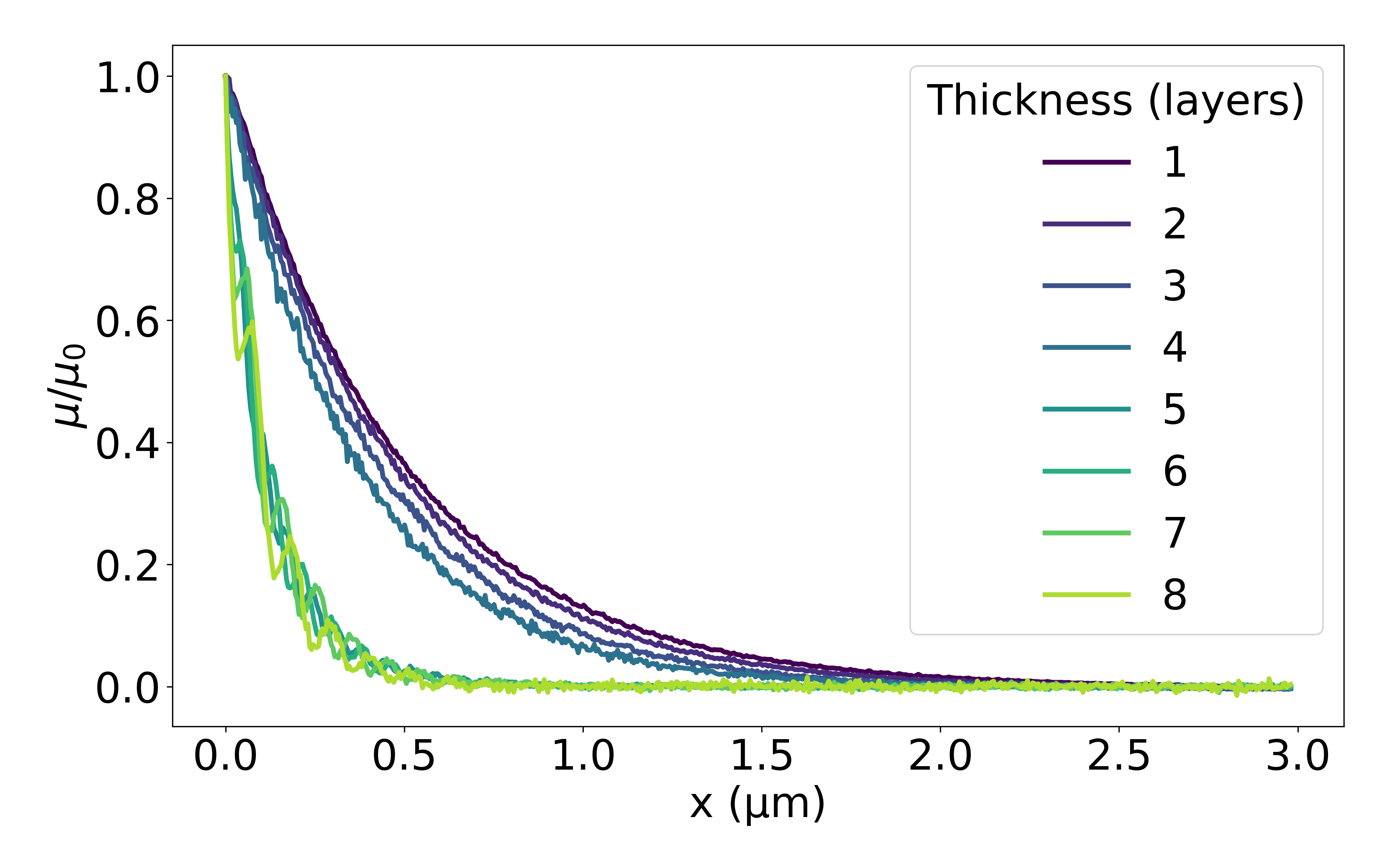}
        \caption{Distance-dependent magnon spin signal in the easy-plane magnetic phase for varying layer thicknesses. The corresponding diffusion lengths are shown in \cref{fig:diffusion_lengths_across_thicknesses}.}
        \label{fig:biaxial_distance-dependent_transport}
    \end{figure}

In \cref{fig:biaxial_distance-dependent_transport}, we show the thickness-dependent magnon spin transport in the easy-plane magnetic phase of hematite in the presence of the critical transverse magnetic field. The magnon diffusion length extracted from these curves is shown with orange triangles in \cref{fig:diffusion_lengths_across_thicknesses}. Similar to the easy-axis magnetic phase, this phase also exhibits a pronounced enhancement in the magnon diffusion length between four and five layers.

\section{Remark on the oscillations} \label{sec:remarkOscillations}
\begin{figure}
        \centering
        \includegraphics[width=\linewidth]{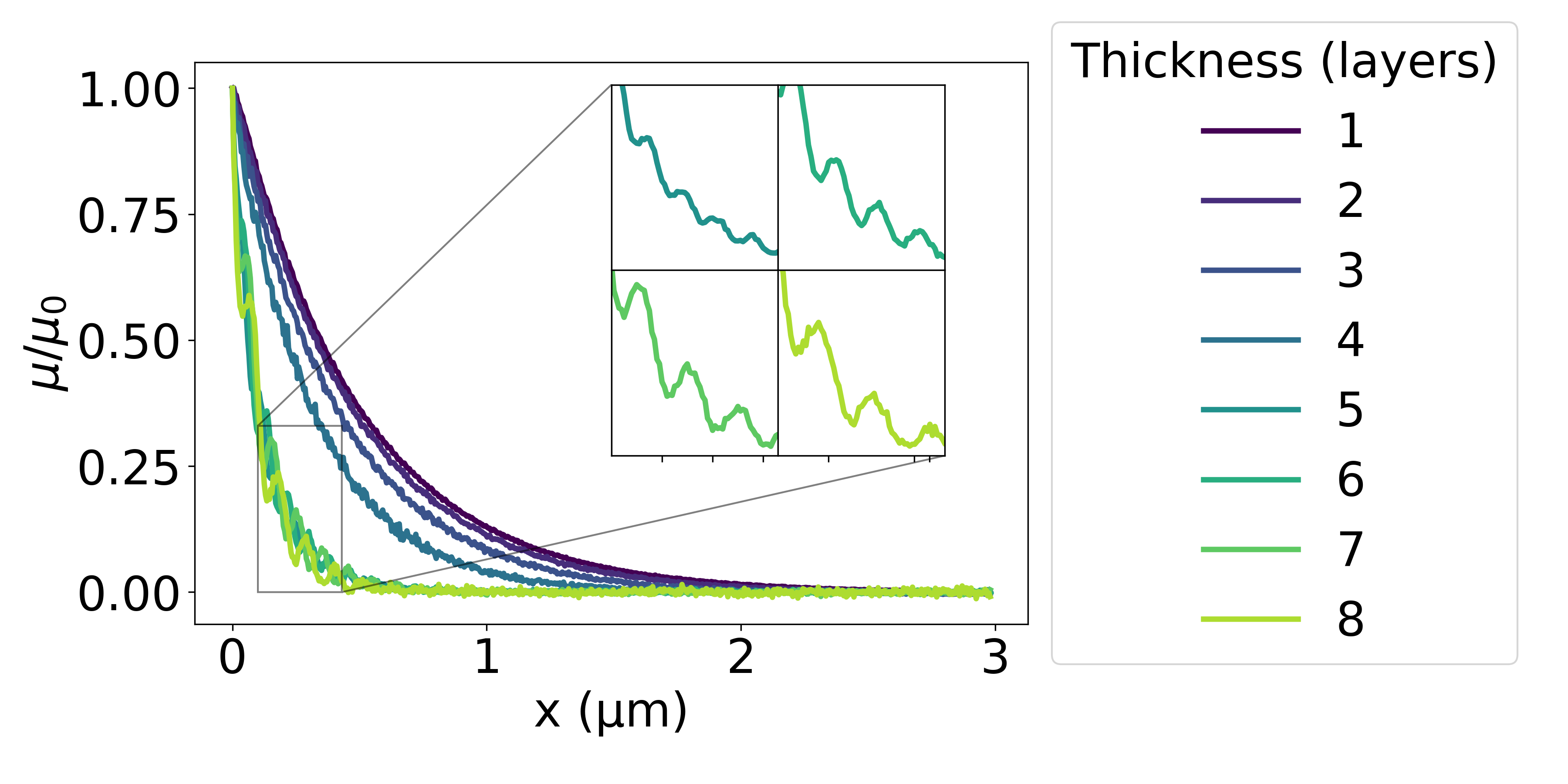}
        \caption{Distance-dependent magnon spin signal in the easy-axis magnetic phase for varying thickness above the crossover thickness. The insets show an oscillation which happens on top of the diffusive transport signal. }
        \label{fig:uniaxial_distance-dependent_transport_with_inset}
    \end{figure}

\begin{figure}
        \centering
        \includegraphics[width=\linewidth]{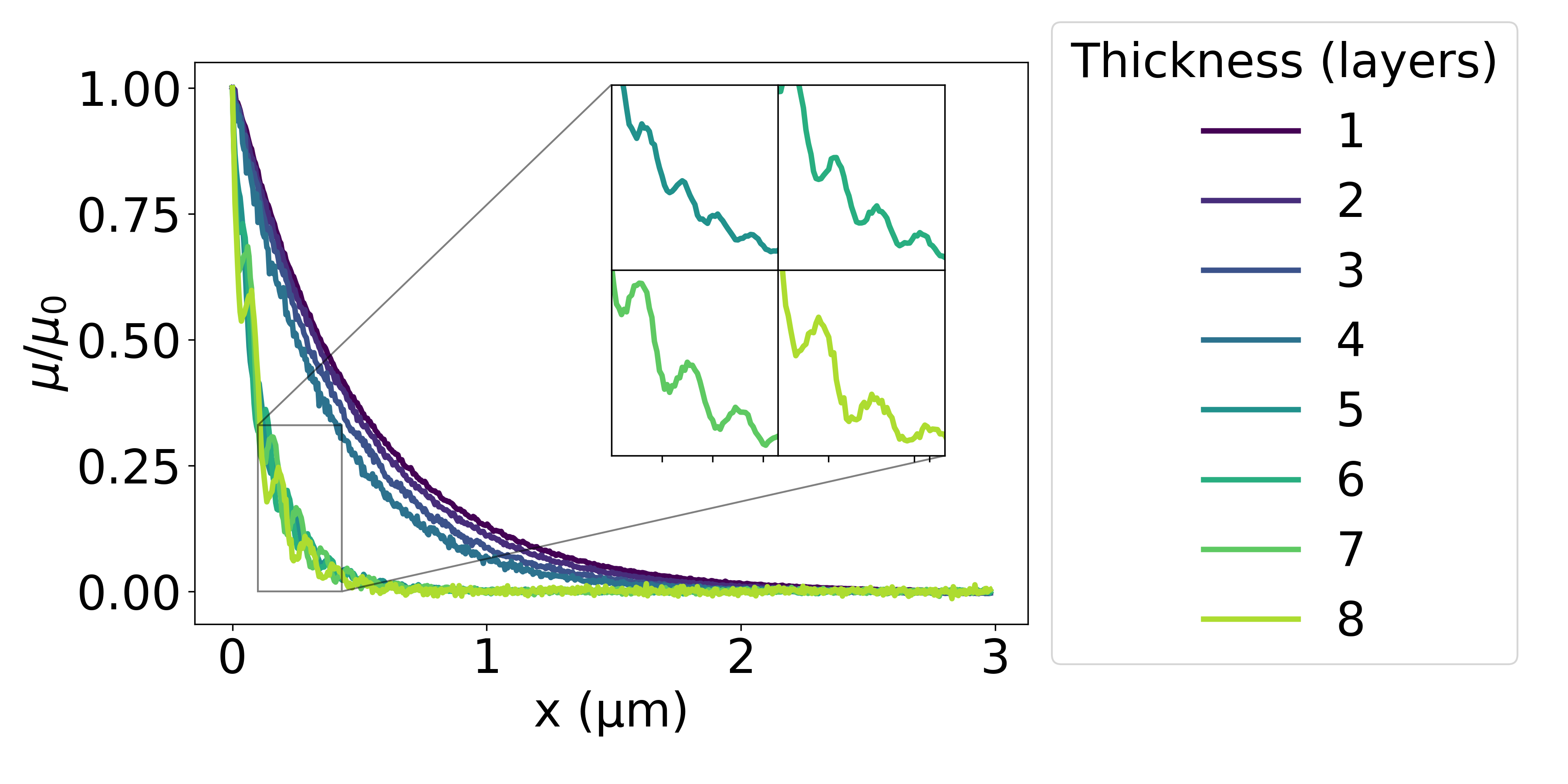}
        \caption{Distance-dependent magnon spin signal in the easy-plane magnetic phase for varying thickness above the crossover thickness. The insets show an oscillation which happens on top of the diffusive transport.}
        \label{fig:biaxial_distance-dependent_transport_with_inset}
    \end{figure}

In the quasi-3D regime, we observe oscillations on top of the exponentially decaying diffusive spin signal; see the insets of
\cref{fig:uniaxial_distance-dependent_transport_with_inset,fig:biaxial_distance-dependent_transport_with_inset}. 
These results suggest that the spin signal in this regime cannot be captured by a single length scale. A model combining two components: a non-oscillating exponential decay, which sets the diffusion length, and an oscillatory decaying term with its own decay length, is required, as given by
\begin{equation}\label{eq:fit_function_harmonic_two_decay_terms}
        \mu^{3D}(x) = \mu_0\left(e^{-\frac{x}{L_D}} + c_2 \sin\left(\frac{x}{\lambda_2}+\phi_2\right) e^{-\frac{x}{l_2}}\right).
    \end{equation}
The exponential decay of the non-oscillatory component of the spin signal is characteristic of diffusive magnon transport, whereas the decaying oscillatory component arises from interference effects due to magnon reflections between the top and bottom interfaces of the thin film. In \cref{tab:fitting_function_two_decays}, we summarize the diffusion length, defined by the decaying signal $L_D$, and the decay length of the oscillatory component $l_2$. Within our material parameters, the latter is about twice as large as the former. The amplitude of the oscillatory component $|c_2|$ is enhanced by increasing the thickness.

    \begin{table}[h]
    \centering
    \begin{tabular}{c c c c}
         \\ \hline \hline 
         Quantity & Symbol & Value & unit \\ 
         \hline
         Length of AFI & $L_x$ & 4 & $\si{\micro\meter}$ \\
         Width of AFI & $L_y$ & 50 & nm \\
         Micromagnetic cell size &  $a$ & 5 & nm \\
         Exchange stiffness & $A$ & 76 & fJ $\text{m}^{-1}$ \\
         Homogeneous exchange & $A_h$ & -460 & kJ $\text{m}^{-3}$ \\
         Easy-axis anisotropy &  $K_{easy}$ & \begin{tabular}{@{}c@{}} -21 (easy-axis phase) \\ -0.021 (easy-plane phase) \end{tabular} &
        \begin{tabular}{@{}c@{}} J $\text{m}^{-3}$ \\ J        $\text{m}^{-3}$ \end{tabular}\\
         Hard-axis anisotropy  & $K_{hard} $ & \begin{tabular}{@{}c@{}} 0 (easy-axis phase) \\ 21 (easy-plane phase) \end{tabular}& 
        \begin{tabular}{@{}c@{}} J $\text{m}^{-3}$ \\ J        $\text{m}^{-3}$ \end{tabular}\\
         Saturation magnetization & $M_s$ & 2.1 & kA $\text{m}^{-1}$ \\
         Gilbert damping & $\alpha$ & $4 \times 10^{-4}$ & - \\
         Time step & $\Delta t$ & 1 & fs \\
         Temperature & $T$ & 0.3 & K \\
         Field-like contribution & $r_g$ & 1 & - \\
         \hline \hline
    \end{tabular} 
    \caption{Parameters used in the stochastic micromagnetic simulations of hematite for both the easy-axis and easy-plane magnetic phases.}
    \label{tab:system_parameters}
    \end{table}

    \begin{table*}%[H]
    %\begin{center}
    %\resizebox{\textwidth}{!}{
    %\setlength{\arrayrulewidth}{0.4pt}
    %\renewcommand{\arraystretch}{1.4}
    \begin{tabular}{|c|c|c|c|c|c|c|c|c|}%{!{\vrule width 1.2pt}P{1.9cm}!{\vrule width 1.2pt}c|c|c|c!{\vrule width 1.2pt}c|c|c|c!{\vrule width 1.2pt}}
    \Xhline{1.2pt}
    \multirow{2}{*}{\shortstack{\textbf{Thickness}\\ \textbf{(layers)}}} & \multicolumn{4}{c!{\vrule width 1.2pt}}{\textbf{Easy-axis phase}} & \multicolumn{4}{c!{\vrule width 1.2pt}}{\textbf{Easy-plane phase}} \\
    \cline{2-9}
    & $L_D$ ($\si{\micro\meter}$) & $l_2$ ($\si{\micro\meter}$) & $|c_2|$ & $\lambda_2$ ($\si{\micro\meter}$)& $L_D$ ($\si{\micro\meter}$) & $l_2$ ($\si{\micro\meter}$) & $|c_2|$ & $\lambda_2$ ($\si{\micro\meter}$)\\
    \Xhline{1.2pt}
    5 & 0.108 & 0.242 & 0.06 & 0.069 & 0.111 & 0.268 & 0.06 & 0.069 \\
    \hline
    6 & 0.110 & 0.260 & 0.103 & 0.081 & 0.109 & 0.256 & 0.101 & 0.081\\
    \hline
    7 & 0.108 & 0.203 & 0.144 & 0.094 & 0.109 & 0.200 & 0.155 & 0.095\\
    \hline
    8 & 0.096 & 0.145 & 0.199 & 0.107 & 0.098 & 0.163 & 0.209 & 0.107\\
    \Xhline{1.2pt}
    \end{tabular}

    \caption{Fitted parameters from Eq.~\ref{eq:fit_function_harmonic_two_decay_terms} for magnon spin signal in the quasi-3D transport regime (5–8 layers), with a constant torque strength of
 $\SI{12}{\micro\volt\per\nano\meter}$. }    \label{tab:fitting_function_two_decays}
    \end{table*}

\section{Impact of system width and spin-torque strength} \label{sec:ImpactWidthAndTorqueStrength}
To further support our findings, we provide additional data showing that the observed crossover thickness does not depend on the thin-film width or the applied spin-torque strength.
 
\subsection{System width dependence of the crossover thickness}
We perform simulations of thickness-dependent spin transport across different system widths $L_y=\SI{5}{nm}$ to $L_y=\SI{1000}{nm}$. For $L_y\leq \SI{50}{nm}$, the critical crossover thickness is between four and five layers, while for $L_y\geq \SI{50}{nm}$, it is between three and four layers for all observed thicknesses. The slightly decreased critical crossover thickness can be understood by the fact that, for wider layers, more transverse modes are available. Therefore, the energy difference between the lowest and higher magnon subbands becomes smaller; thus, occupying higher energy subbands, which leads to quasi-3D transport, is more likely, see \cref{fig:dispersion_8layer}.
    
We conclude that the width of the system does not qualitatively change the quasi-2D to quasi-3D crossover.

\subsection{Spin-torque strength dependence of the crossover thickness}
To investigate the potential dependence of the crossover thickness on spin-torque strength, we systematically vary the torque amplitude, as controlled by the applied voltage, across three distinct cases: two layers, representing the quasi-2D regime; eight layers, corresponding to the 3D regime; and four layers, situated at the crossover thickness. 
For a consistent comparison across the three cases, we maintain a constant relative excitation strength, voltage per thickness, as the spin-torque strength is proportional to the current density, which varies inversely with film thickness; see \cref{eq:torque}.

\begin{figure}
        \centering
        \includegraphics[width=\linewidth]{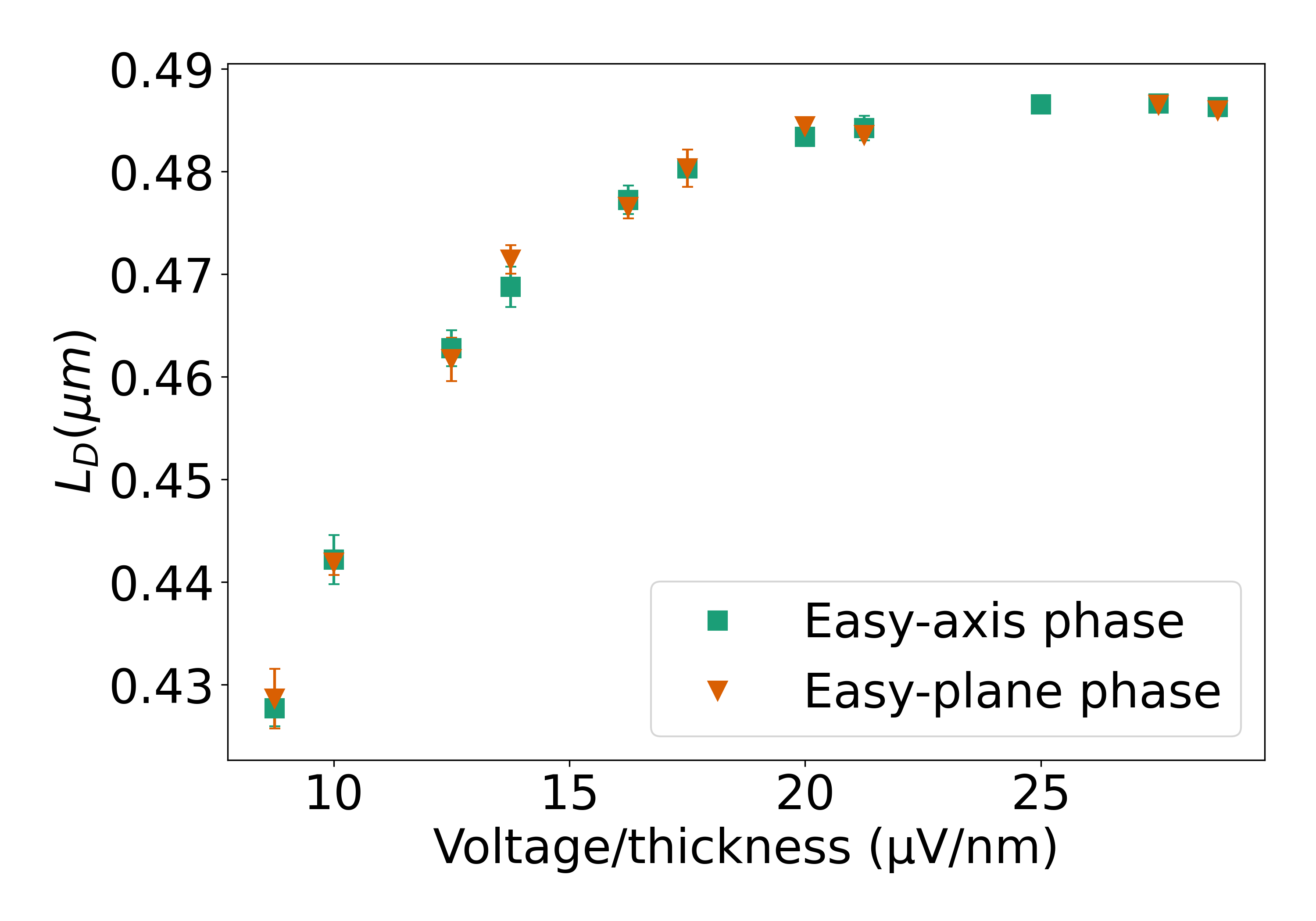}
        \caption{Spin diffusion length for two-layer systems (quasi-2D transport regime) in the easy-axis and easy-plane magnetic phases versus applied voltage or torque strength. Data represent averages over nine ensembles, with standard deviation shown. The diffusion length increases with torque and saturates at intermediate strengths before magnetic reversal at higher strengths.}
        \label{fig:diffusion_length_over_voltage_2layer}
    \end{figure}

As shown in \cref{fig:diffusion_length_over_voltage_2layer}, the diffusion length increases slightly with increasing voltage and saturates around \SI{0.48}{\micro\meter} for both easy-axis and easy-plane magnetic phases. Hence, the magnon spin transport remains within the quasi-2D regime.

\begin{figure}[htbp]
        \centering
        \includegraphics[width=\linewidth]{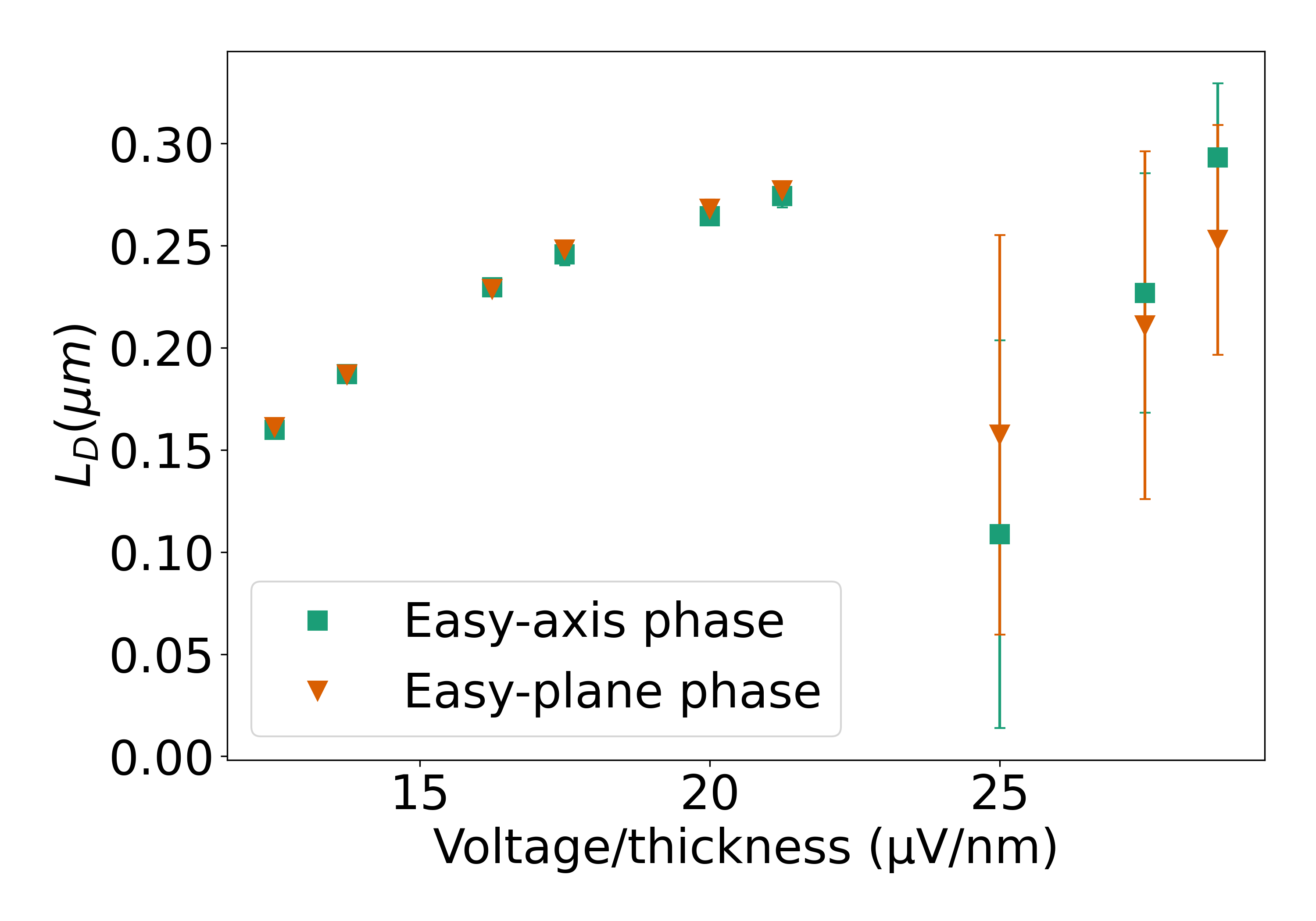}
        \caption{Spin diffusion length for eight-layer thick systems (quasi-3D transport regime)  with varying applied voltage or torque strength in easy-axis and easy-plane magnetic phases.  The diffusion lengths are averages of nine ensembles, and they are plotted together with their standard deviations.}
        \label{fig:diffusion_length_over_voltage_8layer}
    \end{figure}

For the eight-layer system, see \cref{fig:diffusion_length_over_voltage_8layer}, the diffusion length also slightly increases with torque strength but saturates at \SI{0.25}{\micro\meter} before the ground state becomes unstable, as evidenced by large error bars. Again, we conclude that, for both magnetic phases, the magnon spin transport remains well within the quasi-3D regime, independent of the torque strength.

\begin{figure}[H]
        \centering
        \includegraphics[width=\linewidth]{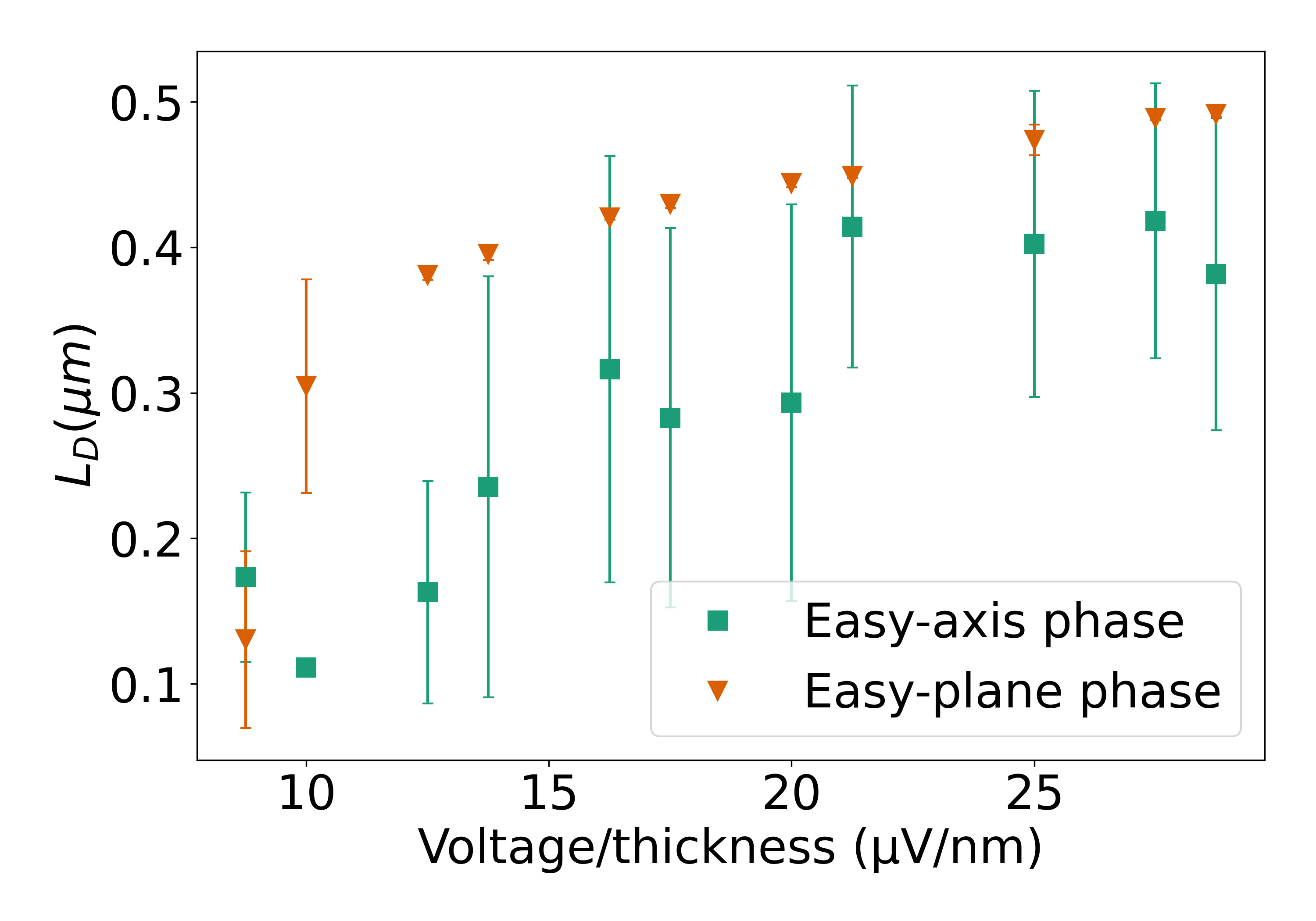}
        \caption{Spin diffusion length for four-layer systems (the crossover thickness) with varying applied voltage or torque strength in two magnetic phases. The diffusion lengths are averages of nine ensembles, and they are plotted together with their standard deviations. The easy-plane magnetic phase shows smaller variance while the easy-axis magnetic phase shows significantly larger standard deviations. }
        \label{fig:diffusion_length_over_voltage_4layer}
    \end{figure}

Finally, right at the crossover thickness, the diffusion length also increases with increasing spin-torque strength, as shown in \cref{fig:diffusion_length_over_voltage_4layer}. However, as indicated by the error bars, the spin signal varies significantly in the easy-axis magnetic phase, and the diffusion length varies from values corresponding to the quasi-2D to the quasi-3D regime. This shows that the crossover occurs around this critical thickness.

\section{Magnon spectra under spin-torque excitations} 
\label{sec:ReadOutOccupiedStates}
\begin{figure}[H]
    \centering
    \includegraphics[width=\linewidth]{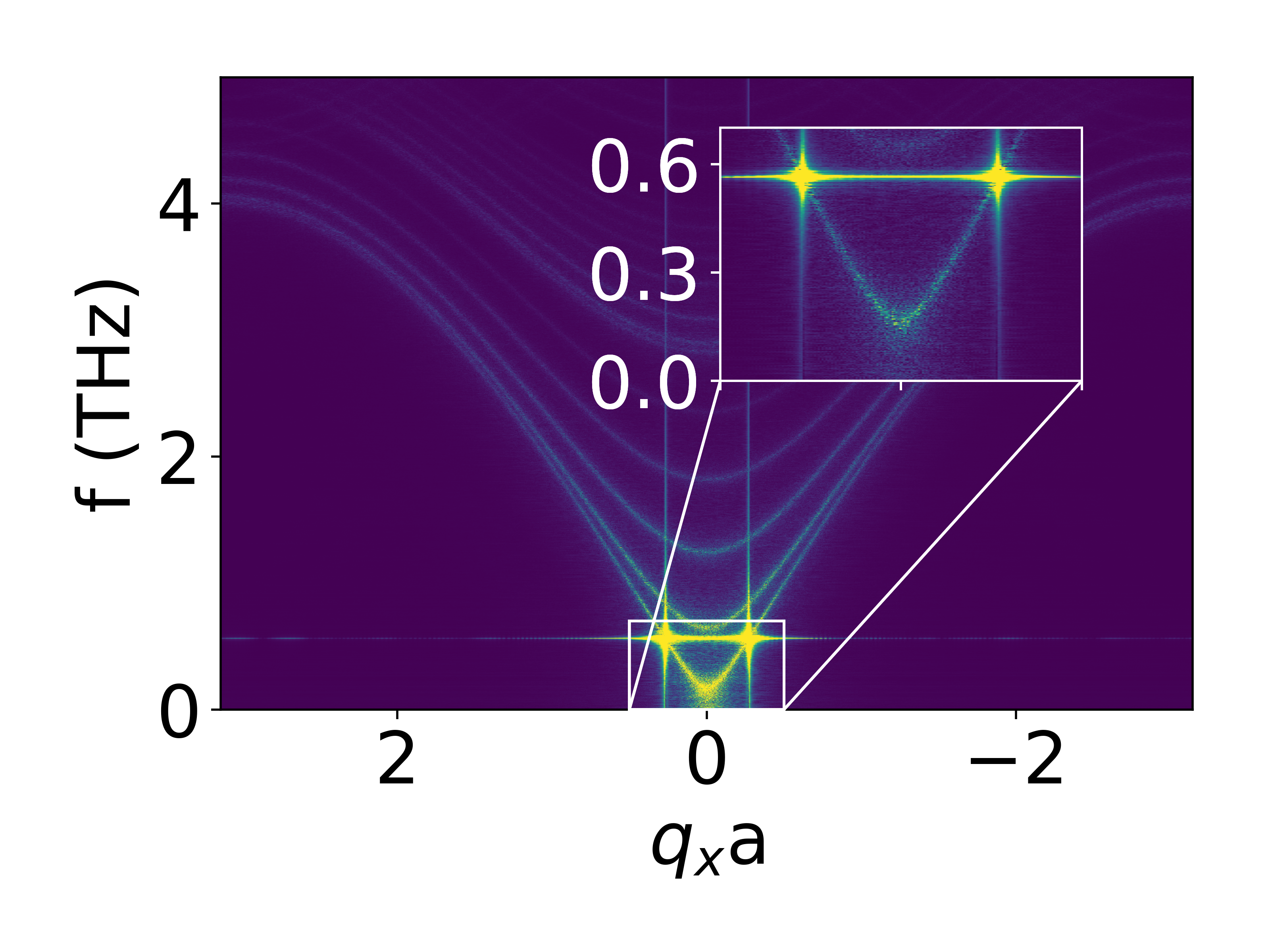}
    \caption{Numerically calculated  magnon dispersion in the easy-axis magnetic phase of a two-layer system (quasi-2D regime) under spin torque with an applied voltage ${|V|}/{L_z} = \SI{19}{\micro\volt \per\nano\meter}$. The excited magnons have a maximum pick at lowest magnon subband in a frequency about $\SI{0.6}{\tera\hertz}$. This leads to a 2D-like magnon spin transport. }
    \label{fig:dispersion_uniaxial_steadystate_2layer_V=-0.188}
    \end{figure}

\begin{figure}[H]
    \centering
    \includegraphics[width=\linewidth]{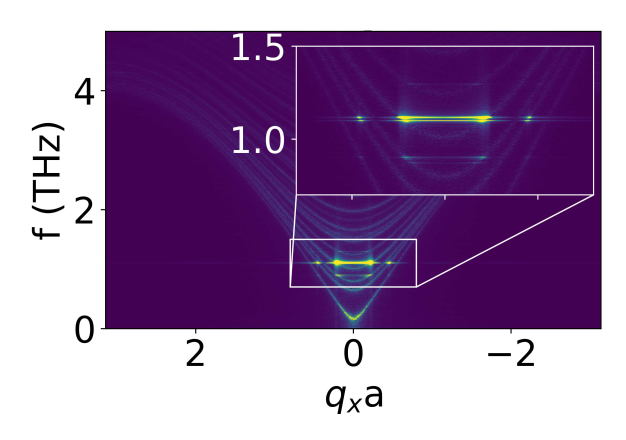}
    \caption{Numerically calculated magnon dispersion in the easy-axis magnetic phase of an eight-layer system (quasi-3D regime) under spin pumping with torque strength of $|V_c|/L_z = \SI{18}{\micro\volt\per\nano\meter}$. Several magnon modes at different frequencies are excited at higher magnon subbands. This leads to a 3D-like magnon spin transport. }
    \label{fig:dispersion_8layer}
    \end{figure}

We record the time-dependent spin configuration during the spin torque excitations and perform a Fourier transform to identify which magnon modes are predominantly excited and contribute to the transport. An example is shown in \cref{fig:dispersion_uniaxial_steadystate_2layer_V=-0.188} for the easy-axis magnetic phase with a thickness of two layers. The color scale represents the Fourier transform amplitude, which is proportional to the number of available magnons, with blue indicating low intensity and yellow indicating high intensity. In our stochastic micromagnetic simulations based on the sLLG equation, all magnon states are thermally populated at low but finite temperatures, following the Rayleigh–Jeans distribution. Superimposed on this thermal background, two prominent signals appear just below \SI{0.6}{THz}, corresponding to magnons injected via spin torque. These states reside in the lowest magnon subband, characterized by $k_y = 0$. The extracted magnon frequencies for all thicknesses are summarized in \cref{fig:critical_frequencies_with_measurements}. On the other hand, above the crossover thickness, higher magnon subbands are mainly excited by the spin torque, indicating a transition from a quasi-2D to a quasi-3D magnon DOS, which results in shorter magnon propagation lengths; see, for example, \cref{fig:dispersion_8layer}. 

This observation can be explained as follows: In thicker films, the reduced energy spacing between quantized modes allows more subbands to lie within the excitation window of spin torque, resulting in the population of additional modes. 
In the quasi-3D DOS, the enlarged phase space for magnon scatterings further increases the probability of scattering events. Consequently, magnons lose coherence and energy more rapidly, leading to shorter propagation lengths.

\begin{figure}
    \centering
    \includegraphics[width=\linewidth]{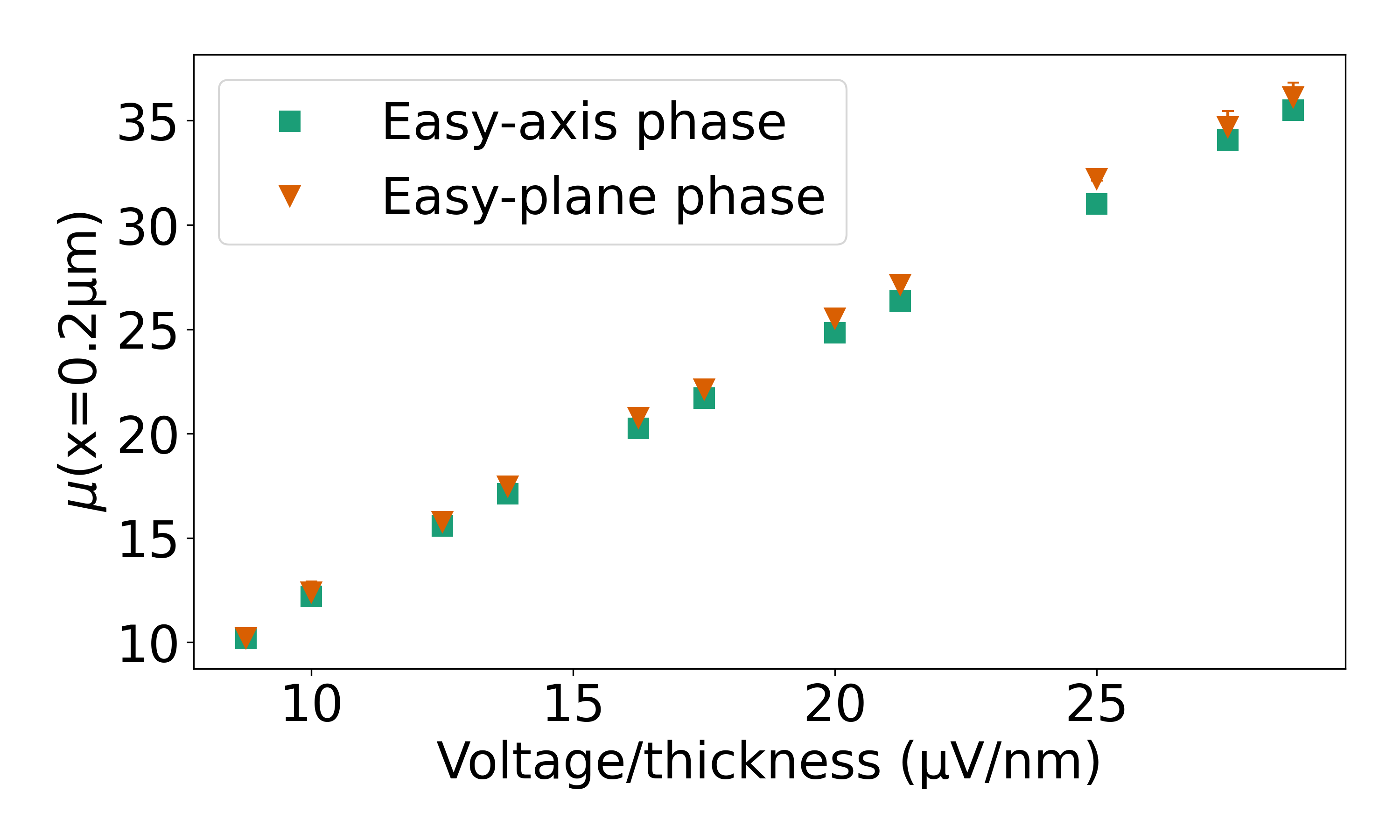}
    \caption{Spin signal $\mu$ at $x=0.2~\mathrm{\mu m}$ as a function of voltage (proportional to the spin torque) for the easy-axis and easy-plane phases in a two-layer system, corresponding to the quasi-2D transport regime. The spin signal exhibits a linear dependence on the applied voltage (or spin torque).} 
    \label{fig:placeholder2layer}
\end{figure}
\begin{figure}
    \centering
    \includegraphics[width=\linewidth]{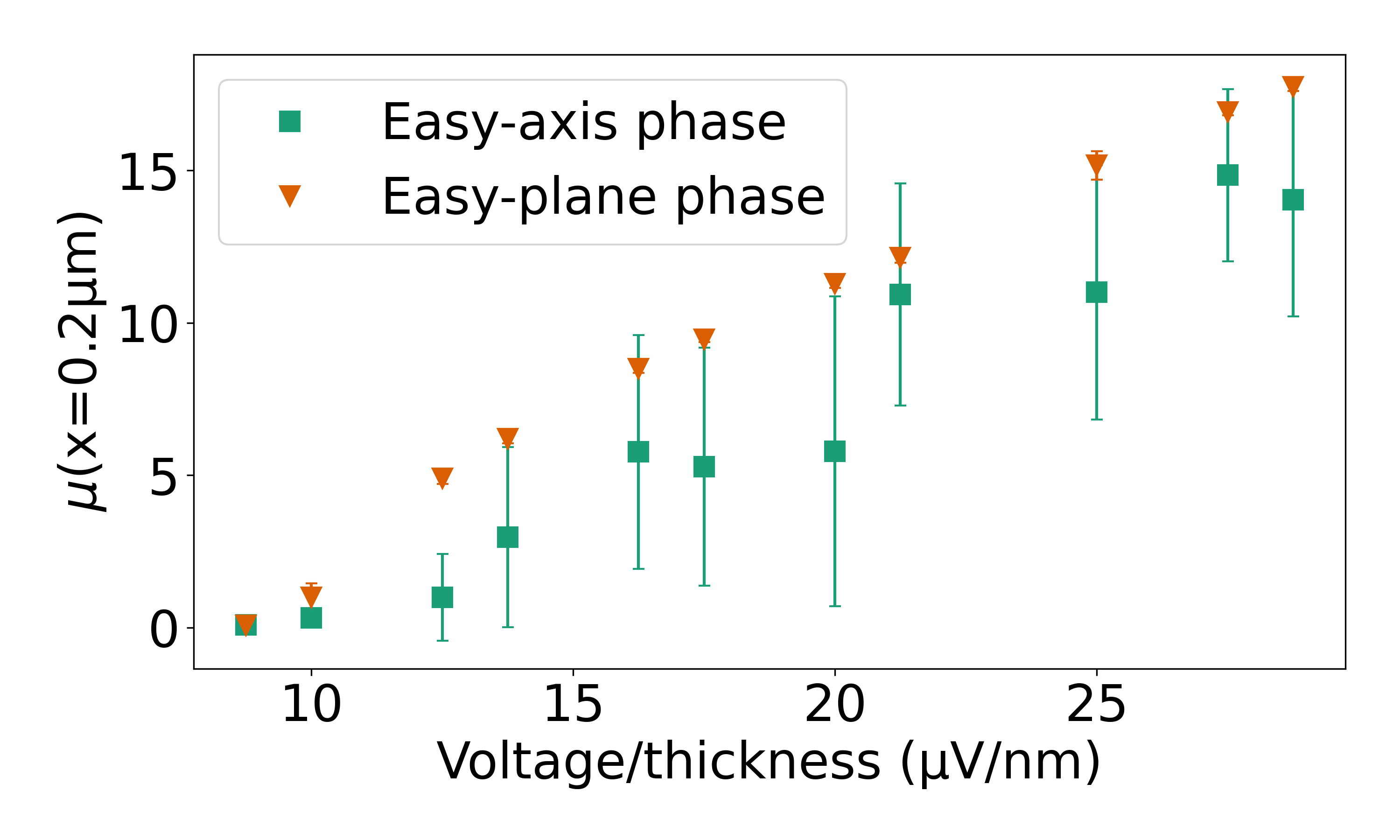}
    \caption{Same as Fig.~\ref{fig:placeholder2layer}, for the four-layer system at the 2D–3D crossover thickness.}
    \label{fig:placeholder4layer}
\end{figure}
\begin{figure}
    \centering
    \includegraphics[width=\linewidth]{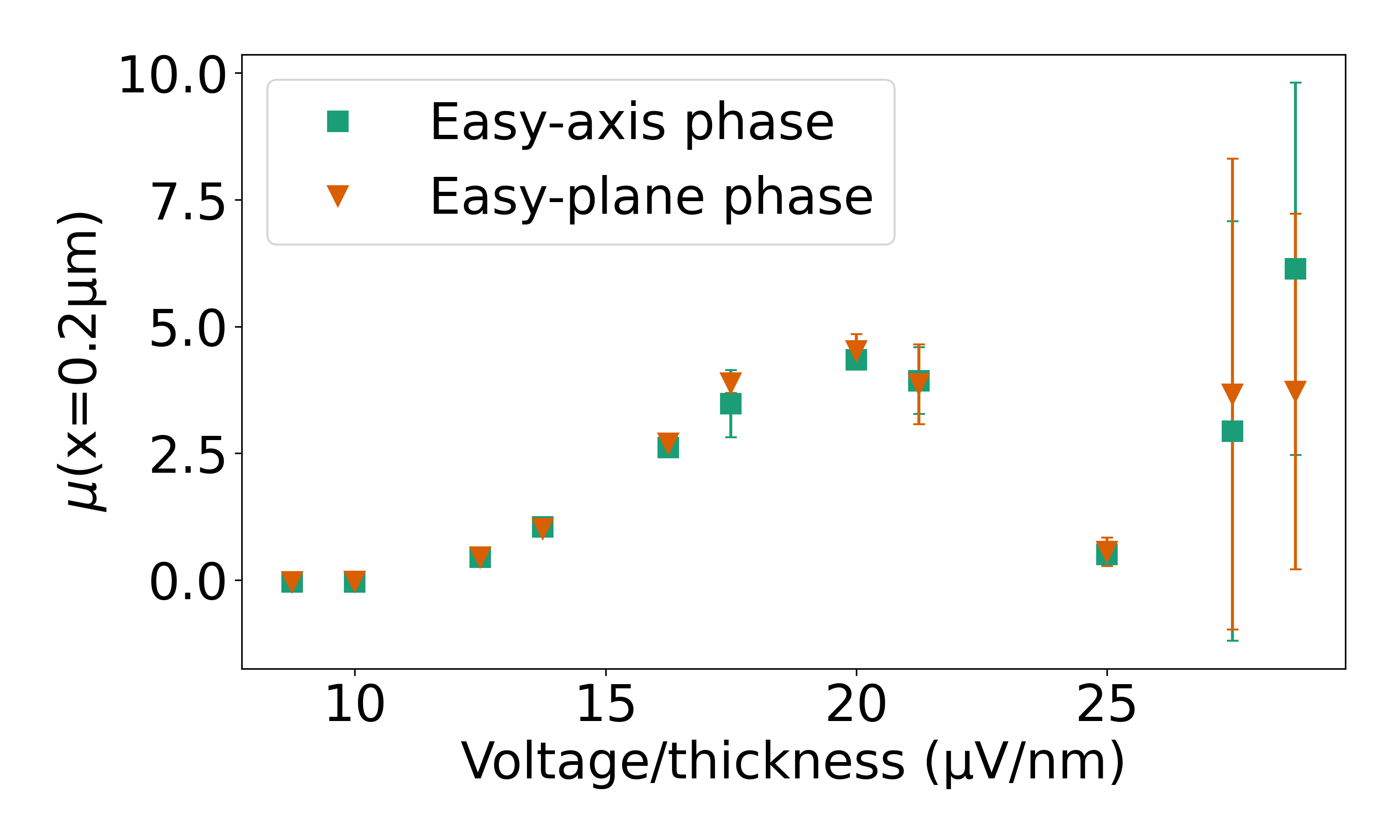}
    \caption{Same as Fig.~\ref{fig:placeholder2layer}, for eight-layer thickness, where the system lies in the quasi-3D transport regime. In this regime the spin signal shows a nonlinear behavior respect to the applied voltage (or the spin torque).}
    \label{fig:placeholder8layer}
\end{figure}

\section{Spin-torque-dependent magnon spin signals} 
In this section, we present the magnon spin accumulation signal at a fixed distance $x=\SI{0.2}{\micro\meter}$ as a function of the applied voltage or spin torque for three transport regimes.
In the quasi-2D regime, the spin accumulation is almost linear with spin-torque strength; see Fig. \ref{fig:placeholder2layer}. However, in the  crossover thickness, Fig. \ref{fig:placeholder4layer}, and quasi-3D regime, Fig. \ref{fig:placeholder8layer}, there are nonlinearities in the magnon spin signal. 

\newpage
\bibliography{bibliography}
\end{document}